\documentclass[reprint,amsmath,amssymb,aps]{revtex4-2}
\usepackage{graphicx}
\usepackage{nicefrac} 
\usepackage{amsfonts}
\usepackage{dcolumn}
\usepackage{bm}
\usepackage[hidelinks]{hyperref}
\usepackage[mathlines]{lineno}
\usepackage{amssymb}
\usepackage{amsmath} 
\usepackage{subfigure}
\usepackage{multirow} 
\usepackage{tabularx} 
\usepackage{array}
\usepackage{units}
\usepackage{tensor} 
\usepackage{braket}
\usepackage{bm}
\usepackage{setspace}
\usepackage{diagbox}
\usepackage{setspace}
\usepackage{xcolor}
\begin{document}
	\title{Investigation of the Effect of Thermal-Induced Atomic Motion on the Conductance of Copper Thin Films}
	\author{Sihe Chen}
	\affiliation{Department of Physics, Binghamton University, Binghamton, New York 13902, USA}
	\author{Kevin Batzinger}
	\affiliation{Department of Physics, Binghamton University, Binghamton, New York 13902, USA}
	\author{Manuel Smeu}
	\email[Corresponding author: ]{msmeu@binghamton.edu}
	\affiliation{Department of Physics, Binghamton University, Binghamton, New York 13902, USA}
	\affiliation{Materials Science and Engineering, Binghamton University, Binghamton, New York 13902, USA}
\begin{abstract}
	Decreases in the size of integrated circuits (IC) and metal interconnects raise resistivity due to the amplification of electron scattering effects, which decreases the efficiency of chiplets. While previous studies have investigated the effect of electron scattering due to a roughened surface, the effect of thermal induced atomic motion on the roughened surface remains unclear. To address this gap, we investigated electron transport in pristine and roughened Cu thin films by performing \textit{ab initio} molecular dynamics (AIMD) trajectories over 20~ps at temperatures of 218~K, 300~K, and 540~K, and then calculating the electron transport properties of the resulting snapshots at 100-fs intervals for the last 10~ps using the non-equilibrium Green's function formalism in combination with density functional theory (NEGF-DFT). As expected, higher temperatures induce larger atomic displacement from equilibrium positions and increase atomic layer separation. We also find that increased temperature results in increased resistance (lower conductance) for the pristine film, but less so for the roughened thin film where the surface roughness itself is the main source of resistance. This study provides insights into how pristine and roughened Cu thin films behave under thermal conditions, helping researchers design better treatments to mitigate thermal effects in ICs and their metal interconnects. 
\end{abstract}
\maketitle
\section{Introduction}
	Copper (Cu) is one of the most fundamental elements for modern microelectronics due to its excellent conductivity and low cost~\cite{guptaCopperInterconnectTechnology2009a,chengCopperMetalSemiconductor2018,besserCurrentChallengesCopper2007}. Since the replacement of aluminum to Cu with interconnects (ICs) in the late 1990s~\cite{kumbhareHighSpeedInterconnectsHistory2022a}, copper has revolutionized the semiconductor industry by enabling higher current densities and improved reliability in interconnect structures. Throughout years of research, a profound understanding of Cu thin films has been developed, from their material properties to their electronic structure to characteristics of resistance~\cite{bisharaApproachesMeasureResistivity2020,anisimovFirstprinciplesCalculationsElectronic1997,valenciaGrainBoundaryResistanceCopper2018,mariniPlanewaveDFTLDACalculation2001,gallSearchMostConductive2020a,mirzaevaComputationalStudyConductance2024}. Much research has been done to understand and improve the conductivity of Cu under various conditions through both theoretical and experimental efforts. One such example is the study by Gan \textit{et al}. which investigates the the conductance of bulk Cu under high pressure and temperature from both theoretical and experimental perspectives~\cite{ganElectricalConductivityCopper2024}. The effort to obtain higher conductance goes beyond pure Cu as well, such as synthesis of various Cu based alloys and graphene-copper composites ~\cite{zhangResearchesHigherElectrical2024,luUltrahighStrengthHigh2004,luStudyHighstrengthHighconductivity2006,panEnhancedElectricalConductivity2022,maassenFirstPrinciplesStudy2010}.  
	
	To meet the demand for higher-performance electronics, companies are seeking to improve their manufacturing processes. One of the most effective ways is to reduce the dimensionality of ICs and increase the packing density of the chip, which also demands the strinking of wiring between IC components. As these ICs shrink, the resistivity of the material increases as it becomes more sensitive to surface and grain boundary scattering~\cite{zahidResistivityThinCu2010,purswaniElectronScatteringSingle2007}. In general, one can describe the surface scattering using the semi-classical Fuchs-Sondheimer (F-S) model~\cite{fuchsConductivityThinMetallic1938,sondheimerMeanFreePath1952}, where the surface specularity \textit{p}, as a phenomenological parameter, describes the scattering at the surface. This parameter ranges from 0 (completely diffuse scattering) to 1 (perfectly specular reflection). However, this parameter does not fully explain the cause of the size effect (increase in resistivity with decreasing dimension) due to the quantum nature of grain-boundaries and rough-surface induced scattering. These factors makes specularity less ideal for theoretical studies \cite{zhouElectricalResistivityRough2018,luoSurfaceRoughnessConductivity1994,rossnagelAlterationCuConductivity2004,timoshevskiiInfluenceSurfaceRoughness2008,schumacherNewEvidenceValidity1987,zhouResistivityScalingDue2018a,munozSizeEffectsCharge2017,huSizeEffectResistivity2023}. 
	
	Another challenge is the self-heating effect which creates a positive feedback loop, where resistivity induces more Joule heating which further increases resistivity through lattice distortion, inducing signal distortions and impedance in current flow~\cite{plombonInfluencePhononGeometry2006}. Experimental work has been done to investigate the temperature dependence of resistivity in bulk and nanowire Cu~\cite{josellSizeDependentResistivityNanoscale2009,matulaElectricalResistivityCopper1979}. Gaining insights into how thermal motion and surface roughness under various temperatures affect the electron transport properties is an important direction for research. From a fabrication perspective, thermal atomic layer etching has been proposed as a potential strategy to mitigate the impact of surface roughness by enabling more controlled surface smoothing~\cite{georgeProspectsThermalAtomic2016,fischerThermalAtomicLayer2021,gerritsenSurfaceSmoothingAtomic2022}. However, a degree of surface roughness is still necessary for reliable bonding to barrier layers. Research by others has investigated properties in roughened Cu thin film surfaces, such as how the resistivity changes with the thickness of Cu thin films with one- and two-sided roughness. It was found that surface roughness, as well as the thickness of the film, affects the conductance of the Cu thin film~\cite{keResistivityThinCu2009a}. Additionally, one recent study has elucidated how bias voltage can somewhat overcome the effect of a defect in Cu thin films. This study also showed a trend where barrier layer atoms implanted on the surface exhibit highest conductance when the barrier layer atom is in the same group on the periodic table as Cu~\cite{batzingerComputationalInvestigationElectron2024c}. However, these works focused on 0~K structures, while the effect from distortions induced by a thermal environment remains unclear. The purpose of this study is to provide such insights and understand the temperature effect on the conductance of Cu by analyzing structures generated from molecular dynamics trajectories and calculating electron transport properties of these structures. 
\section{Computational Methods}
	Structural optimization calculations were performed \textit{via} density functional theory (DFT); atomic motion and quantum transport properties were simulated using \textit{ab initio} molecular dynamics (AIMD) and the non-equilibrium Green's function technique(NEGF), respectively~\cite{dattaQuantumTransportAtom2012a}. DFT and AIMD~\cite{kresseInitioMolecularDynamics1993,carUnifiedApproachMolecular1985} calculations were carried out using the Vienna \textit{ab initio} Simulation Package (VASP)~\cite{kresseEfficientIterativeSchemes1996}. The package uses plane wave basis sets with projector augmented wave (PAW) potentials~\cite{blochlProjectorAugmentedwaveMethod1994a} and the Perdew–Burke–Ernzerhof exchange–correlation functional (PBE)~\cite{perdewGeneralizedGradientApproximation1996b}. All AIMD trajectories used a $\Gamma$-centered 1$\times$1$\times1$ Monkhorst-Pack~\cite{monkhorstSpecialPointsBrillouinzone1976} \textit{k}-point grid and Methfessel-Paxton~\cite{methfesselHighprecisionSamplingBrillouinzone1989} smearing along with a 500-eV plane-wave cutoff. The primitive cell for the Cu thin film was obtained from the Materials Project repository (MP-30)~\cite{jainCommentaryMaterialsProject2013}. Cell manipulation and trajectory analysis were done using Ovito~\cite{stukowskiVisualizationAnalysisAtomistic2010} and VESTA~\cite{mommaVESTA3Threedimensional2011}. The electron transport properties were calculated \textit{via} the NEGF technique in conjunction with DFT, as implemented in NanoDCAL~\cite{taylorInitioModelingQuantum2001}, an electron transport simulation package that starts by constructing the Hamiltonian of the simulation cell through first principles, then utilizes the Keldysh NEGF formalism to obtain the transmission function $T(E)$, which is the probability that an electron with a given energy $E$ will successfully transmit through a structure. Lastly, when an electron is scattered in the central region, we can construct a scattering matrix (or $S$-matrix) to represent this event. The $S$-matrix contains transmission and reflection coefficients that can be useful in identifying changes in electron flow through the eigenchannels of the electrodes which inject and extract electrons to and from the central region. The derivation of the transmission function and the scattering matrix is given in Section~S1 of the Supplemental Material~\cite{SM}.
	\subsection{Simulation Cell}
\begin{figure}[!h]
			\hspace*{-1em}
			\includegraphics[scale=0.28]{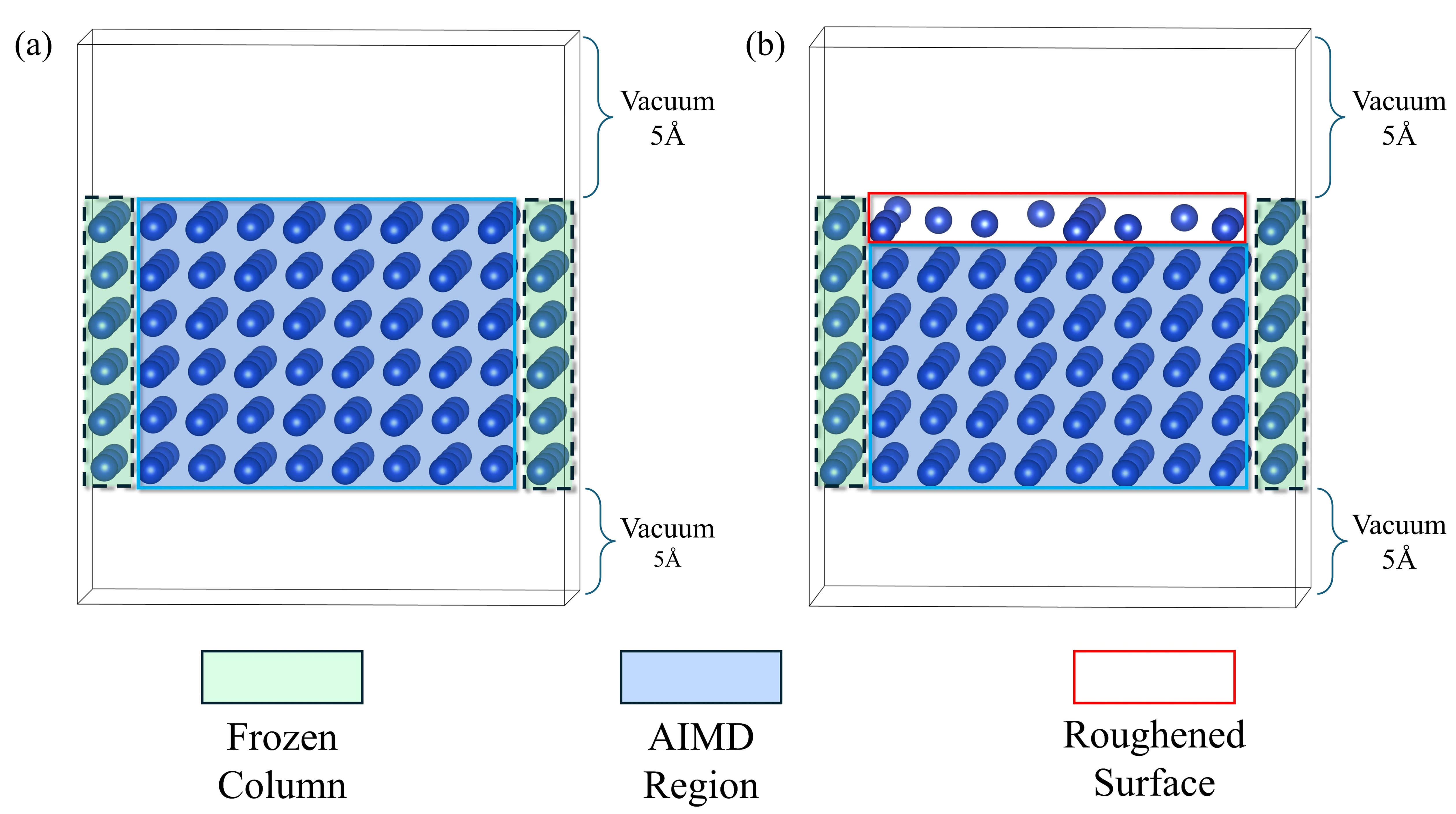}
			\caption{Illustration of the simulation cell. Panel (a) shows the pristine simulation cell with 5~\AA~of vacuum on top and bottom of the Cu slab with frozen regions on the left and right of the cell. Panel (b) shows the simulation cell with the roughened top surface.}
			\label{Cell Illustration}
		\end{figure}

		For the pristine Cu system, we first performed a structural relaxation of the fcc unit cell. The self-consistent field (SCF) loop was converged to an energy difference of $1 \times 10^{-6}$~eV, and the interatomic forces were converged to $<$~0.02~eV/\AA. Each Cu atom had 11 valence electrons, while the remaining 18 electrons were treated within the pseudopotential. The calculated lattice constant (3.63~\AA) was in good agreement with the experimental value (3.61~\AA) \cite{kuriyamaXrayDiffractionCrystal1969}. With the relaxed fcc Cu structure, we then expanded it into a 3{$\times$}3{$\times$}5 supercell to construct the AIMD cell. The cell was then extended by 5~{\AA} of vacuum at the top and bottom to simulate the effect of a thin film, as shown in Fig.~\ref{Cell Illustration}. 
		Details and a flowchart for the cell construction procedure are provided in Fig.~S1 of the Supplemental Material~\cite{SM}. The system contained a total of six atomic layers with 30 Cu atoms in each layer. For the roughened surface copper, we modified the pristine structure by removing Cu atoms from the topmost layer of the structure. A total of three percentages of roughness were considered: 4~\%, 50~\% and 100~\%, which correspond to removing 1, 12 and 24 Cu atoms, respectively. The leftmost and rightmost layers of atoms were frozen to maintain a stable connection between the electrodes and the central region, and therefore are not considered part of the surface layer. Note that the surface atoms of the frozen regions did not count towards the overall number of Cu atoms on the surface.
	\subsection{Interlayer Spacing \& Simulation Parameters}
		To account for interlayer spacing between the system due to thermal expansion, we performed a linear temperature ramp from 1~K to the desired temperature followed by equilibration. During the ramping stage, we used the Nosé-Hoover thermostat~\cite{evansNoseHooverThermostat1985a} and increased the temperature from 1~to 540~K at a rate of 1~K/fs. After reaching temperatures of 218~K, 300~K and 540~K, which correspond to low-, room- and high-temperature operation, we extracted the generated structures and performed an additional 10-ps equilibration with the NVT ensemble. The {\textit{y}}-coordinates were recorded over the last 5~ps and averaged to obtain the interlayer spacing. Examples of the \textit{y} positions of all Cu atoms at 300~K over the 10-ps equilibration are provided in Fig.~S2 of the Supplemental Material~\cite{SM}. 
	
		With the interlayer spacing adjusted for all three temperatures, we started new AIMD trajectories with the NVT ensemble. These trajectories were run for 20~ps with 1-fs time steps, where the first 10~ps were utilized for equilibration and the last 10~ps were considered production runs. During production runs, structures were extracted every 100~fs and utilized for quantum transport calculations. These procedures were also applied to roughened surface Cu systems. Note that the 50~\% roughened system was run with three temperatures, while the other two roughness percentages (4~\% and 100~\%) were only performed at 300~K. Due to limited computational resources, we focused on the extreme roughness values (0~\% and 50~\%) at three temperatures. We found that temperature causes a larger reduction in transmission of the pristine system than in the 50~\% roughness system; therefore, we expect the trends for 4~\% and 100~\% roughness to lie between those of the two extremes.
\section{Results and Discussion}
	\subsection{Interlayer Spacing}
		From the initial AIMD trajectories, layer expansion was observed. This effect is expected due to thermal expansion, and is more prominent in higher temperature trajectories. The values of interlayer spacing expansion is summarized in Table~\ref{Interlayer Spacing}, which shows an increase in interlayer spacing with higher temperature.
		\begin{table}
			\caption{Interlayer spacing (\AA) for the Cu film.}
			\label{Interlayer Spacing}
			\begin{ruledtabular}
				\begin{tabular}{c c c c c}
					\diagbox{Layers}{\shortstack{Temperature\\(K)}} & 0 & 218 & 300 & 540 \\
					\hline  \\
					Layer 1 to 2 & 1.76 & 1.80 & 1.81 & 1.88 \\
					Layer 2 to 3 & 1.82 & 1.79 & 1.81 & 1.87 \\
					Layer 3 to 4 & 1.82 & 1.81 & 1.82 & 1.90 \\
					Layer 4 to 5 & 1.82 & 1.80 & 1.81 & 1.87 \\
					Layer 5 to 6 & 1.75 & 1.80 & 1.81 & 1.88 \\
				\end{tabular}
			\end{ruledtabular}
		\end{table}
	\subsection{Electron Transport Properties}
		\subsubsection{Pristine Cu Slab}
			Tranmission spectra of the trajectories were sampled every 100~fs over the last 10~ps of the trajectory. These transmission spectra were compared to those for the initial structure of the AIMD trajectories. This initial structure is the configuration used to start the AIMD run: all Cu atoms are at their fcc crystallographic sites. For a given temperature, the coordinates of Cu at each layer are uniformly adjusted so that the interlayer spacings match the values listed in Table~\ref{Interlayer Spacing}. Thus, the initial structure represents an ordered, thermally expanded lattice with each Cu layer at its crystallographic position and the appropriate temperature-dependent interlayer spacing.
			
\begin{figure}[!h]
				\centering
				\includegraphics[scale=0.5]{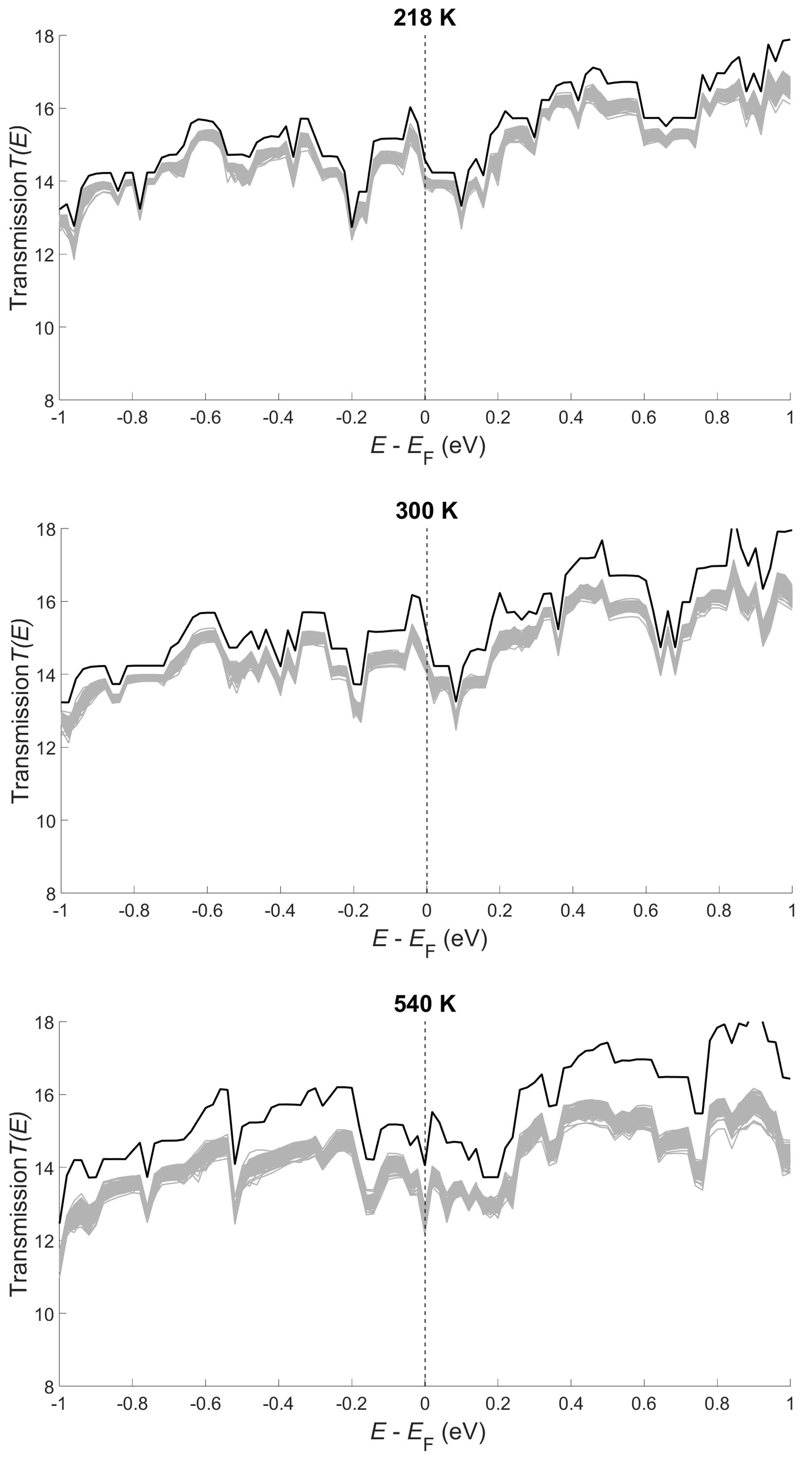}
				\caption{Transmission spectra for the pristine Cu slab at 218~K, 300~K and 540~K. The black series are the initial structures with the adjusted interlayer spacing for that particular temperature, and the gray series are snapshots taken from AIMD trajectories at their corresponding temperatures.}
				\label{Pristine Transmission}
			\end{figure}

			For the dynamics of the Cu slab under a thermal environment, we performed AIMD trajectories and extracted structures from these trajectories and define them as the AIMD structures. Information regarding the trajectories used for the electron transport calculations is provided in Fig.~S3 of the Supplemental Material~\cite{SM}, including their instantaneous energy and temperature. Due to the thermal environment, Cu atoms were displaced from their crystallographic position, which induced lattice distortion. Using the structures generated from the initial structures and AIMD trajectories, we performed electron transport calculations and obtained their transmission spectra $T(E)$ . The obtained transmission spectra are shown in Fig.~\ref{Pristine Transmission}. At each temperature, thermal motion lowers the transmission (gray) when compared to their respective initial structure (black). As expected, the reduction effect is greater at higher temperature. The transmission values at the Fermi level are shown in Table~S1 of the Supplemental Material~\cite{SM}.
			
			Because of the steep slope of $T(E)$ near the Fermi level ($E_\mathrm{F}$), the transmission value at $E_\mathrm{F}$ is sensitive to small shifts in energy. To account for this sensitivity, we estimated the low-bias conductance, $G$, by integrating $T(E)$ near the Fermi level with a small bias window, 
			\begin{equation}
				G=\frac{1}{0.2}\int_{-0.1}^{0.1}T(E)dE\,
			\label{Conductance Equation}
			\end{equation}
			where $G$ denotes the low-bias conductance, $0.2$ denotes the integration window and -0.1 to 0.1 are energetic bounds of integration. This calculation was performed in two ways: For the initial structure, there was only one conductance value at each temperature; For the AIMD structures (218~K, 300~K and 540~K), the low-bias conductance was averaged over the set of snapshots. At 0~K there was no AIMD trajectory, so only the initial structure conductance is given. These results are summarized in Table~\ref{Pristine Conductance Value}. For the initial structures, the conductance is relatively similar despite having different interlayer spacings, indicating that if the Cu atoms are kept in their crystallographic positions, the spacing between layers does not play a significant role in the conductance. However, when introducing thermal motion into the system using AIMD, we observed an inverse relationship between conductance and temperature, where an increase in temperature decreases the conductance. On the other hand, standard deviations in conductance show a direct relationship but remain small. This indicates that despite greater atomic displacement induced by higher temperature throughout trajectories, variations between each snapshot have minimal effects on the overall averaged low-bias conductance. 
			\begin{table}
				\caption{Low-bias conductance of the initial and AIMD averaged pristine Cu thin film structures.}
				\label{Pristine Conductance Value}
				\begin{ruledtabular}
					\begin{tabular}{c c c}
						\shortstack{Temperature\\(K)} & \shortstack{Initial Structures\\($G_\mathrm{0}$)} & \shortstack{AIMD Structures $\pm$ $\sigma$\\($G_\mathrm{0}$)}\\
						\hline
						 0  & 14.68 &                  \\
						218 & 14.77 & 14.29~$\pm$ 0.06 \\
						300 & 14.85 & 14.11~$\pm$ 0.08 \\
						540 & 14.89 & 13.54~$\pm$ 0.11 \\
					\end{tabular}
				\end{ruledtabular}
			\end{table}
		\subsubsection{Roughened Cu Surface}
			After introducing roughness to the surface of the system, a similar procedure was performed to determine the electron transport properties of the roughened surface systems. Due to limited computational resources, only a single roughness profile was examined in this case. However, we direct interested readers to our previous work~\cite{batzingerComputationalInvestigationElectron2024c}, where varying roughness configurations are compared for a smaller and static system. We started by inducing different surface roughness percentages to the pristine thin film structure and the top-down view of these roughness profiles are shown in Fig.~S4 of the Supplemental Material~\cite{SM}. Figure~\ref{Percentage_Roughness}a shows the transmission of three different percentages of roughness as well as the pristine system (0~\%). When inspecting the transmission spectra of 0\% (complete surface) and 4\% surface roughness (1 surface vacancy), a minor decrease is expected due to electron scattering at the vacancy. Next, a significant reduction in transmission was observed between 4\% and 50\% (12 surface vacancies) due to the increase in the number of scattering sites and a large reduction in the number of Cu atoms. However, when comparing the transmission between 50\% and 100\% (24 surface vacancies), an increase in transmission was observed despite having fewer Cu atoms in the system. This phenomenon indicates that an increase in surface roughness up to 50\% drastically lowers the transmission, but further increases in the surface roughness percentage beyond 50\% allow the system to partially recover some of the transmission lost. This trend translates to the low-bias conductance of the different surface roughness systems at 0~K: There is a 2.85\% reduction between 0\% and 4\% surface roughness; a 28.97\% reduction between 4\% and 50\% surface roughness; and an 18.24\% increase between 50\% and 100\% surface roughness.
			\begin{figure}[!h]
				\centering
				\includegraphics[scale=0.45]{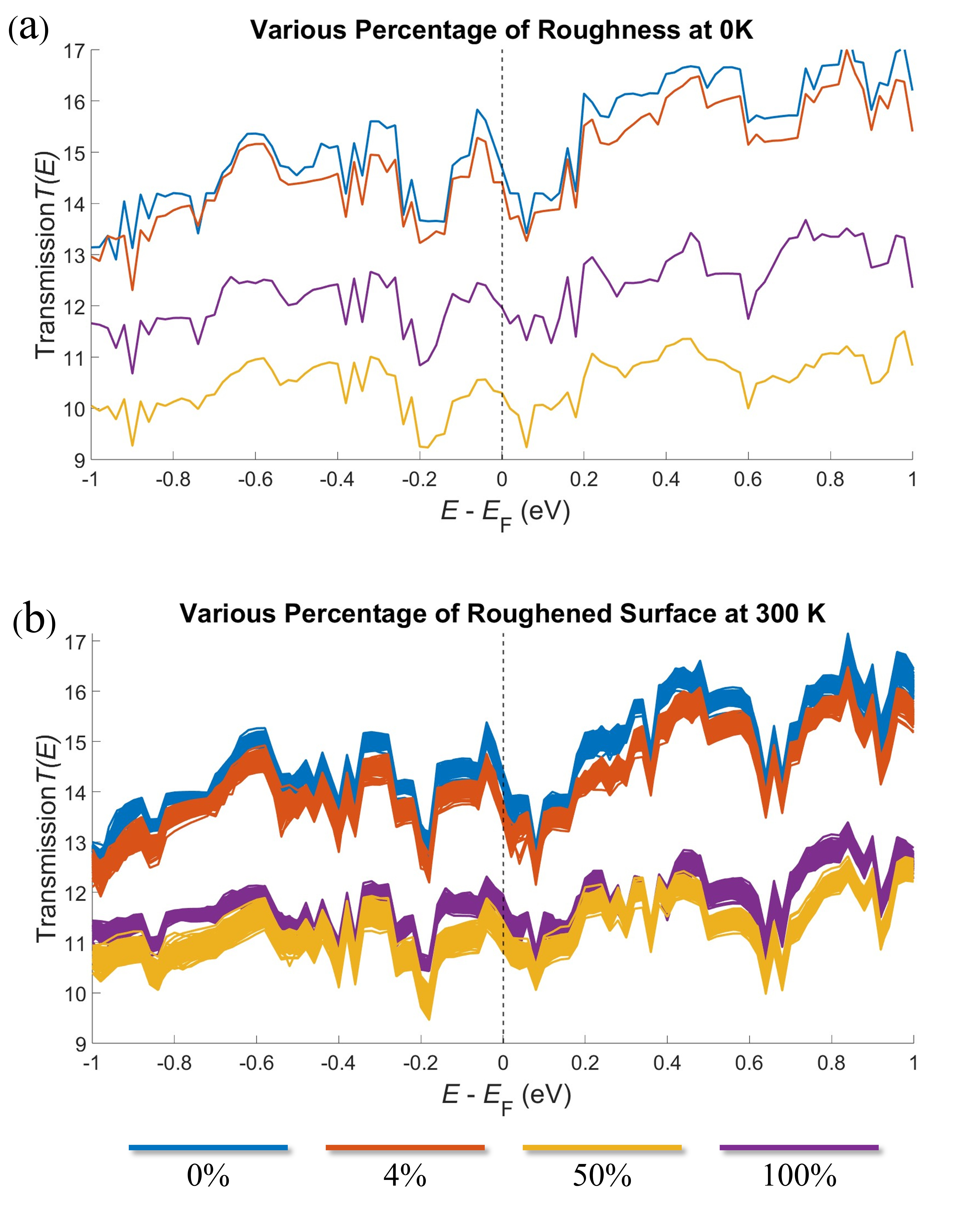}
				\caption{Transmission spectra for Cu thin films with 0~\% (pristine), 4~\%, 50~\% and 100~\% surface roughness at (a) 0~K and (b) 300~K.}
				\label{Percentage_Roughness}
			\end{figure}
			
			When the roughened surface systems are exposed to the 300~K thermal environment, a trend similar to the static (0~K) system was observed, shown in Fig.~\ref{Percentage_Roughness}b. Comparing the 100~\% and 50~\% systems, we noticed that the 100~\% transmission spectra are, again, higher than 50~\%. This phenomenon can be explained in terms of the surface morphology. From 0~\% to 50~\% roughness, the number of vacancies and scattering sites increases, which reduces the conductance. For roughness above 50~\%, the surface can instead be viewed as a thinner slab with adatoms on top rather than a thicker slab with a higher number of vacancies. This morphology contains fewer adatoms, and thereby fewer scattering sites, leading to higher conductance. Values of low-bias conductance for the various surface roughness percentages are summarized in Table~\ref{Percentage_Roughness_Table}. Results for both temperatures (0~K and 300~K) agrees with the results of Ke \textit{et al.}~\cite{keResistivityThinCu2009a}, which showed an inverse parabolic behavior when varying the percentage of surface roughness, that resistivity of a surface with concentration $x$ is very close to the resistivity of a surface with concentration of $(1-x)$ and maximized at 50~\%. This suggests that surface integrity can play a critical role in optimizing conductance and this property translates to Cu thin film even at higher temperature. 
			\begin{table}
				\caption{Low-bias conductance of Cu thin films with various percentages of surface roughness.}
				\label{Percentage_Roughness_Table}
				\begin{ruledtabular}
					\begin{tabular}{c c c}
						\shortstack{Roughness\\Percentage\\(\%)} & \shortstack{Conductance\\at 0~K\\($G_\mathrm{0}$)} & \shortstack{Averaged\\Conductance at 300~K\\($G_\mathrm{0}$)}\\
						\hline
						0   & 14.68 & 14.11~$\pm$ 0.06 \\
						4   & 14.25 & 13.60~$\pm$ 0.08 \\
						50  & 10.13 & 10.98~$\pm$ 0.11 \\
						100 & 11.96 & 11.58~$\pm$ 0.07 \\
					\end{tabular}
				\end{ruledtabular}
			\end{table}
			
			To further investigate the effect of thermal induced motion on a roughened surface, we performed additional AIMD simulation under different temperatures. The transmission spectra of the 50~\% roughened systems (Fig.~S4~\cite{SM}) at different temperatures are shown in Fig.~\ref{Defected Transmission} and transmission values at the Fermi level are shown in Section S4B of the Supplemental Material~\cite{SM}. Previous work has shown that the maximum conductance reduction occurs at this percentage~\cite{keResistivityThinCu2009a}. From the figure, one immediate observation is that transmission spectrum of the initial structure (black series) becomes nearly indistinguishable from the set of transmission spectra of the AIMD structures (gray series). This indicates that the distortion introduced by the thermally-induced atomic displacement no longer affects the electron transport properties greatly. Such distortion introduced by the temperature will have a minimal effect if the surface integrity of the Cu thin film is already compromised. We also observed that some of the AIMD structures have transmission spectra values slightly higher than that of the initial structure due to favorable scattering. This shows that when surface integrity is compromised, thermally-induced distortions can actually aid electron transport and help restore some of the conductance of the Cu thin film.
			\begin{figure}
				\includegraphics[scale=0.5]{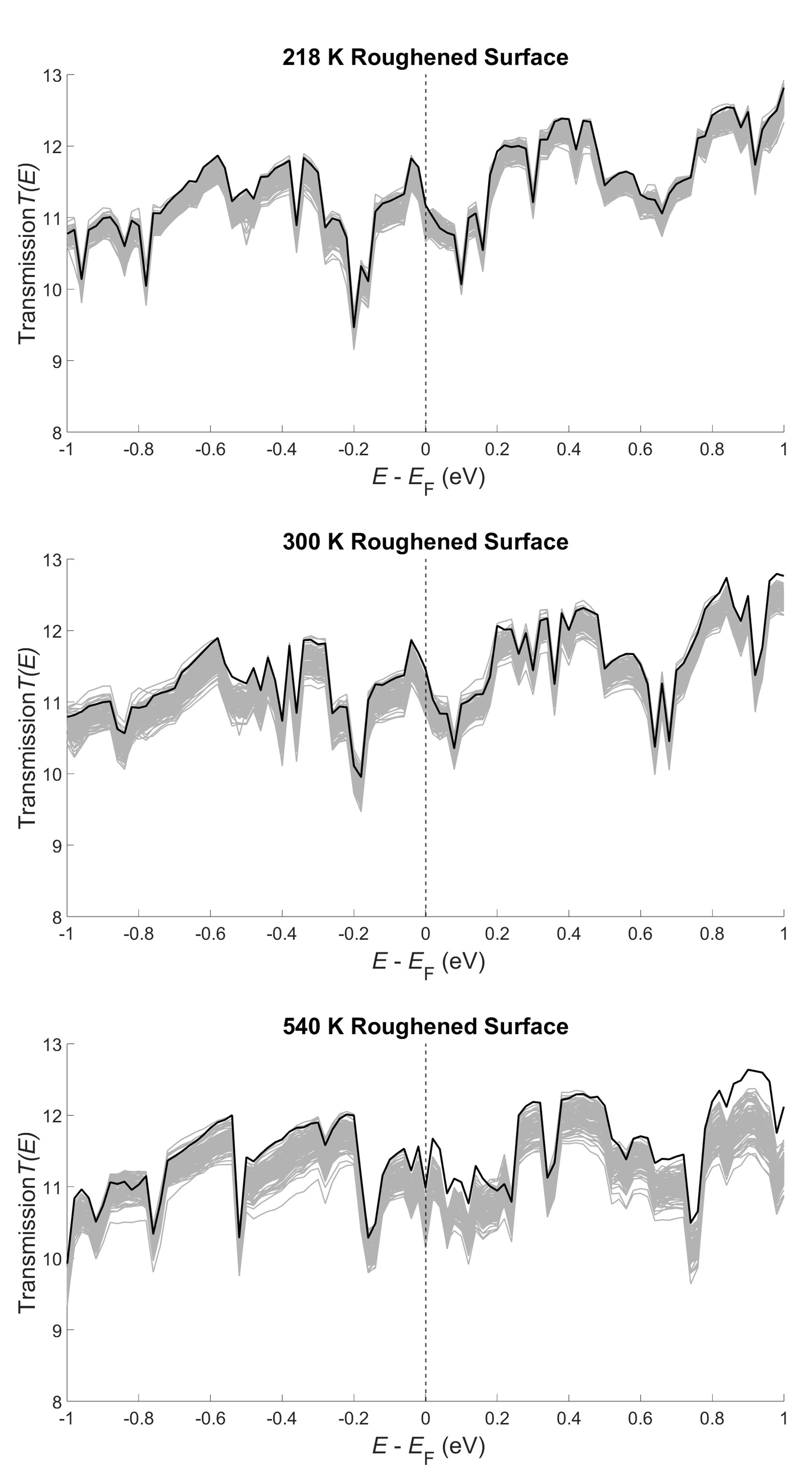}
				\centering
				\caption{Transmission spectra for the Cu slab with 50~\% surface roughness at 218~K, 300~K and 540~K. The black series are the initial structures and the gray series are AIMD snapshots.}
				\label{Defected Transmission}
			\end{figure}
			 The AIMD averaged low-bias conductances of the Cu thin film with 50~\% roughness at 218~K, 300~K and 540~K as well as that of the initial structure are summarized in Table~\ref{Roughened_Conductance_Value}. One intriguing finding is that the conductance of the initial structures of the 50~\% roughened surface increases at higher temperature. This can be understood from the increased interlayer spacing described in Table~\ref{Interlayer Spacing}, particularly between the roughened layer and the second layer (layers 1 and 2). As the separation increases, the roughened layer influences transport through the underlying layers to a lesser extent.
				\begin{table}
				\caption{Low-bias conductance of the 50~\% roughened surface.}
				\label{Roughened_Conductance_Value}
				\begin{ruledtabular}
					\begin{tabular}{c c c}
						\shortstack{Temperature\\(K)} & \shortstack{Initial Structures\\($G_\mathrm{0}$)} & \shortstack{AIMD Structures $\pm$ $\sigma$\\($G_\mathrm{0}$)}\\
						\hline
						 0  & 10.13 &                  \\
						218 & 11.14 & 11.00~$\pm$ 0.09 \\
						300 & 11.19 & 10.98~$\pm$ 0.11 \\
						540 & 11.32 & 10.88~$\pm$ 0.12 \\
					\end{tabular}
				\end{ruledtabular}
				\end{table}  
			
			Note that when comparing the initial and final structure of the 50~\% roughened Cu thin film, we observed that a total of three Cu atoms migrated from their initial position at 540~K. These Cu atoms migrated during the early stages of the AIMD trajectories (equilibration period), but did not migrate further during the production run afterward. These migrations are further discussed in Section S6 of the Supplemental Material~\cite{SM}. No other atom migration occurred in any of the other AIMD trajectories. 
		\subsubsection{Summary}
\begin{figure}[!h]
				\includegraphics[scale=0.27]{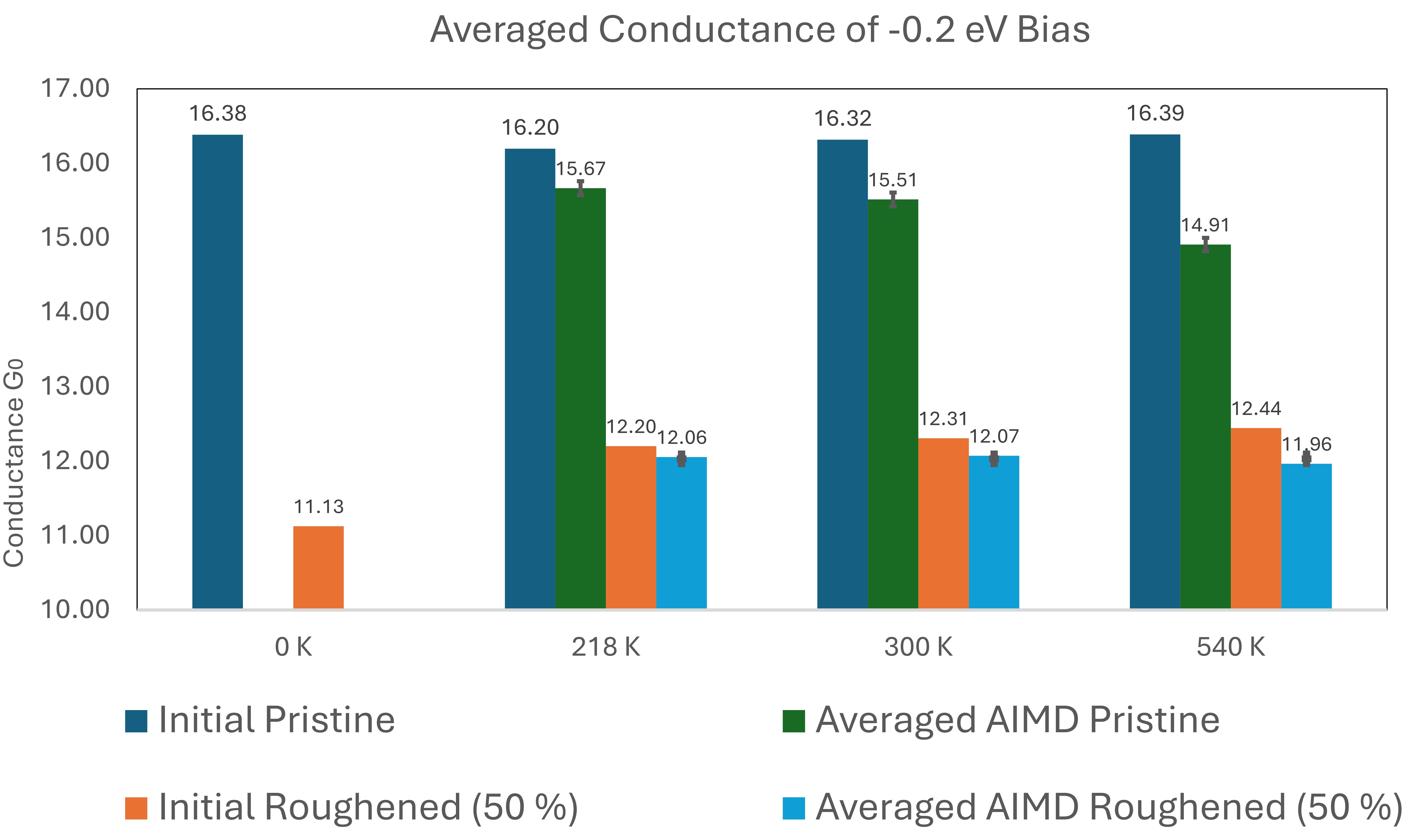}
				\caption{Overall low-bias conductance of the pristine and 50~\% surface roughness systems. Standard deviations are indicated on top of each bar.}
				\label{All Conductance}
			\end{figure}

		Figure~\ref{All Conductance} shows all the calculated low-bias conductance values, including the pristine systems and 50~\% surface roughness systems at the three temperatures.
		Comparison of the low-bias conductance values of the pristine systems (dark blue bars) reveals similar values despite having different interlayer spacings. This suggests that slight adjustment to the interlayer spacing does not affect the electron transport properties of Cu appreciably. After introducing the temperature factor \textit{via} AIMD trajectories, we observed different levels of conductance reduction in the averaged AIMD pristine system (green bars). As we increased the temperature, the reduction from initial pristine system to averaged AIMD pristine system (dark blue and green bars) became larger since Cu atoms displaced farther from their equilibrium positions. For the 50~\% roughened system, we observed a drastic reduction in the conductance when comparing the 0~K roughened system (orange bars) with the 0~K pristine system (dark blue bars). However, unlike the pristine system, when comparing the roughened AIMD-average with their corresponding initial roughened systems (light blue and orange bars), the reduction was much smaller. This indicates that distortion induced by thermal motion does not play a significant role in the conductance when surface integrity is already compromised.
		\subsubsection{Scattering-State Analysis}
		We performed a scattering-state analysis for the pristine system and the system with 50~\% surface roughness at 300~K, as shown in Fig.~\ref{Scattering_States}. The explanation of the scattering-state matrix is provided in Section~S1 of the Supplemental Material~\cite{SM}. For each AIMD snapshot, we extracted the transmission coefficients for each eigenchannel of the scattering matrix and computed the channel-resolved transmission coefficient by summing the magnitudes (moduli) of the relevant transmission matrix elements. It should be pointed out that we only performed this analysis for one of the transverse \textit{k}-points while the transmission and conductance results described in the preceding sections correspond to averages over the four transverse \textit{k}-points; therefore a direct comparison is not expected. These values were then plotted at each timestep used for the transport calculations (once every 100 fs). In total, 17 transmission channels are available. We highlight three channels with distinctive behaviors, while the remaining channels are shown in Fig.~S6 of the Supplemental Material~\cite{SM}.
		
		The pristine system shows near maximum transmission coefficients in each channel for the initial structure, which are slightly lowered due to thermal motion at 300~K (blue dashed and solid lines in Fig.~\ref{Scattering_States}, respectively). Introducing 50~\% surface roughness dramatically affects some of the transmission channels (e.g., Channel 3 and 7, among others shown in Fig.~S6) for the initial structure (dashed orange lines), yet other channels are not as strongly affected (e.g., Channel 13). Most interestingly, once atomic motion is taking place, we see that some of the reduced transmission can be partially recovered at some snapshots, as shown for Channel 7 in Fig.~\ref{Scattering_States}.
			\begin{figure}[!h]
				\includegraphics[scale=0.5]{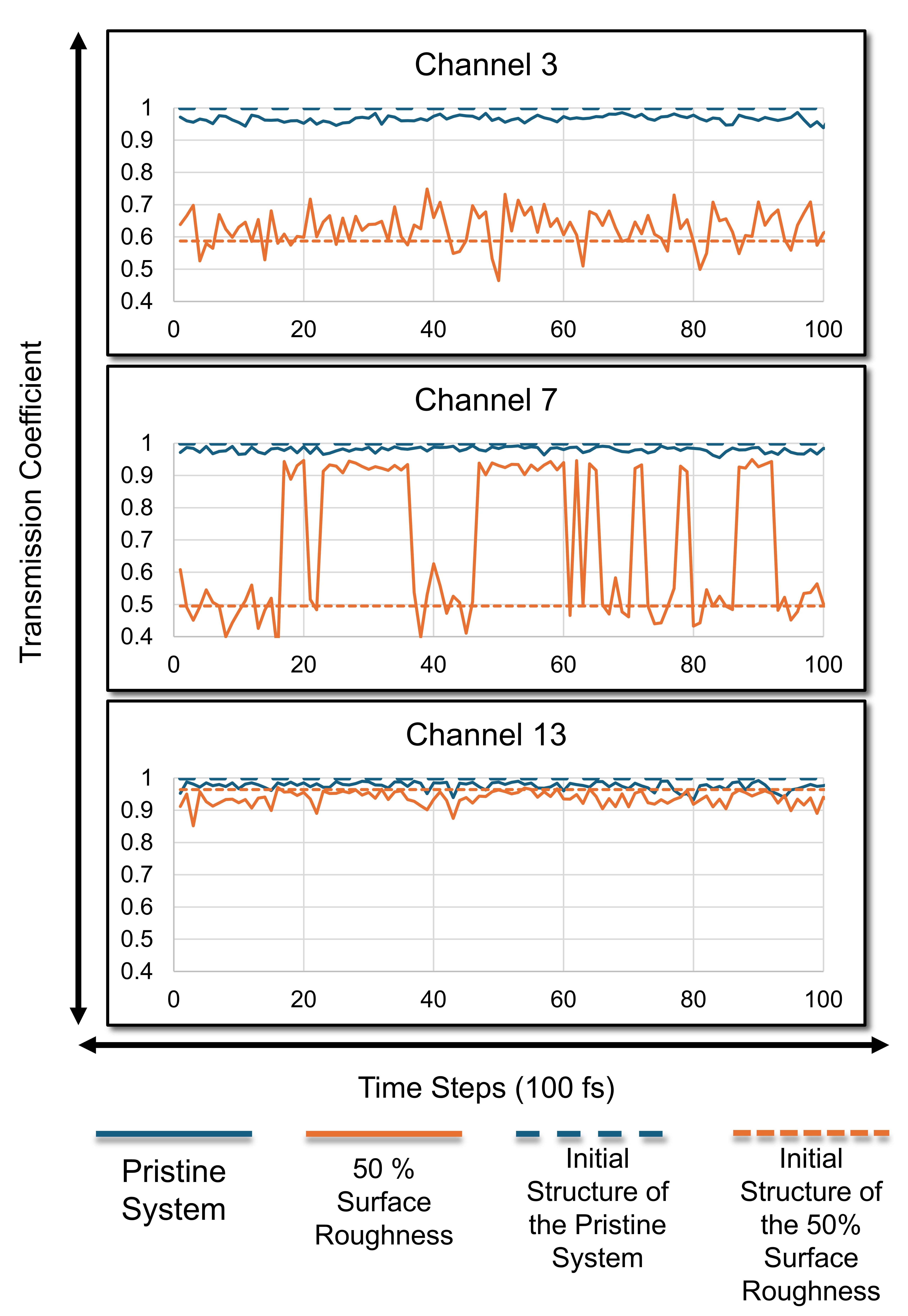}
				\caption{Channel-resolved transmission coefficients for select channels as a function of time (solid lines), compared to the values for the initial structure (dashed lines) for the pristine (blue) and 50~\% surface roughness (orange) systems.}
				\label{Scattering_States}
			\end{figure}
\subsection{Conclusion}
	From the discussion above, we have shown how the electron transport properties such as transmission spectra and low-bias conductance of Cu thin films changes when accounting for thermal atomic motion with AIMD. For the pristine system, there is a clear reduction in conductance proportional to the temperature, associated with increased atomic motion, which is demonstrated through the transmission spectra and the low-bias conductance. Roughening the surface will cause a significant penalty to the conductance, with thermally-induced structural distortions playing a much smaller role in reducing the conductance. This study further clarifies how pristine and roughened copper (Cu) surfaces conduct electrons across different temperatures and highlights the importance of surface integrity.
\subsection{Acknowledgment}
	We acknowledge financial support from the Integrated Electronics Engineering Center (IEEC) at Binghamton University as well as Binghamton University's High Performance Computing (HPC) cluster, {\textit{Spiedie}}, for computational resources.
\newpage
\bibliography{Reference/Citation}
\end{document}